\documentclass[%
 reprint,
 amsmath,amssymb,
 aps,
]{revtex4-1}
\usepackage{graphicx}
\usepackage{dcolumn}
\usepackage{bm}
\usepackage{graphicx}
\usepackage{dcolumn}
\usepackage{bm}
\usepackage{bigints}
\usepackage{color}
\usepackage{amsmath,amssymb,graphicx}
\usepackage{float}
\usepackage{braket}
\usepackage{comment}
\usepackage[normalem]{ulem}

\def\lpzo{$\rm LP_{01}$}
\def\lpzt{$\rm LP_{02}$}

\def\st{$\mathcal{S}$}
\def\as{$\mathcal{A}$}
\begin{document}

\preprint{APS/123-QED}

\title{Apparent non-conservation of momentum of light due to strongly coupled nonlinear dynamics in a multimode optical fiber}

\author{Hamed Pourbeyram}
\author{Arash Mafi}
\email{mafi@unm.edu}
\affiliation{Center for High Technology Materials, Univ. of New Mexico, Albuquerque, NM 87106, USA}%
\affiliation{Department of Physics \& Astronomy, Univ. of New Mexico, Albuquerque, NM 87131, USA}%

\date{\today}

\begin{abstract}
The intimate interaction between phase-matched parametric amplification and strong nonlinear mode coupling
in a multimode optical fiber results in a saturable spatial mode conversion that appears to violate
the conservation of momentum. We investigate, by theory and experiment, this new regime of strong nonlinear 
interaction in multimode optical fibers and show that the appearance of the non-conservation of linear 
momentum is merely a clever sleight of hand by nature. This novel saturable mode conversion can potentially 
be used in a variety of applications including the generation of exotic quantum states of light.
\end{abstract}

\pacs{Valid PACS appear here}
\maketitle
Multimode optical fibers (MMFs) combine the rich dispersive and nonlinear properties~\cite{1974-Stolen,1981-Hill,1982-Stolen-Bjorkholm,1981-Lin} of light propagation in the controlled 
space of a few optical modes to develop a greater mastery of the spectral and spatial properties of 
light~\cite{2016-Christodoulides-LopezGalmiche,2016-Krupa-PRL,2015-Wise-Wright-PRL,2012-Mecozzi,2016-Wen,2016-Toulouse-Esmaeelpour,2016-Wise-Zhu,2013-Mafi-Hamed,2007-Xu-Lee,2016-Kurpa-Wabnitz,2015-Peyghambarian,2013-Nazemosadat-Mafi,2013-Essiambre,2015-Hamed-photonics,2005-Ramachandran,2012-Xu-Cheng}.
Even with only a few key dispersive and nonlinear processes governing the propagation of light in 
single-mode optical fibers, their interplay results in such vast and rich dynamics that has been the subject of intense research 
over the past few decades~\cite{2013-Agrawal-Book}. The presence of multiple spatial modes in a MMF elevates the complexity and
richness of the possibilities to a whole new level~\cite{2008-Poletti,2012-Mafi,2015-Wise-wright,2016-Sivan,2015-Picozzi,2013-Pedersen}. 
The added complexity in MMFs can sometimes result is anomalous observations and unexpected results. 
A prominent example is the yet unexplained report on the violation of angular momentum conservation in 1981 by 
Hill et al.~\cite{1981-Hill}. Recently, we carried out a series of experiments in similar settings to that of Hill et al.~\cite{1981-Hill} to investigate the
intermodal (IM) amplified spontaneous four-wave mixing (FWM)--we observed what appeared to be a violation of the 
conservation of linear momentum. This Letter is an attempt to explain this and similar anomalous observations in experiments involving MMFs.  
Needless to say that the conservation laws stand as expected and the observed non-conservation of linear momentum is merely a clever 
sleight of hand by nature.

\begin{figure}[t]
\includegraphics[width=3.2in]{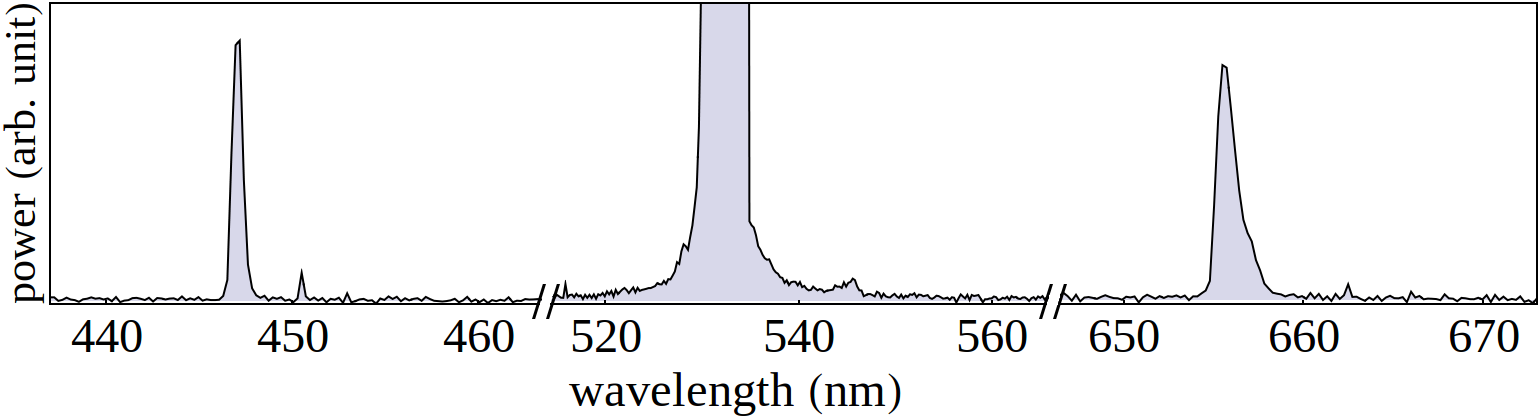}
\caption{Observed spectrum for the nonlinear process discussed in the paper, i.e. strongly coupled FWM and NLC.}
\label{Fig:spectrum}
\end{figure}

\begin{figure}[t]
\includegraphics[width=3.2in]{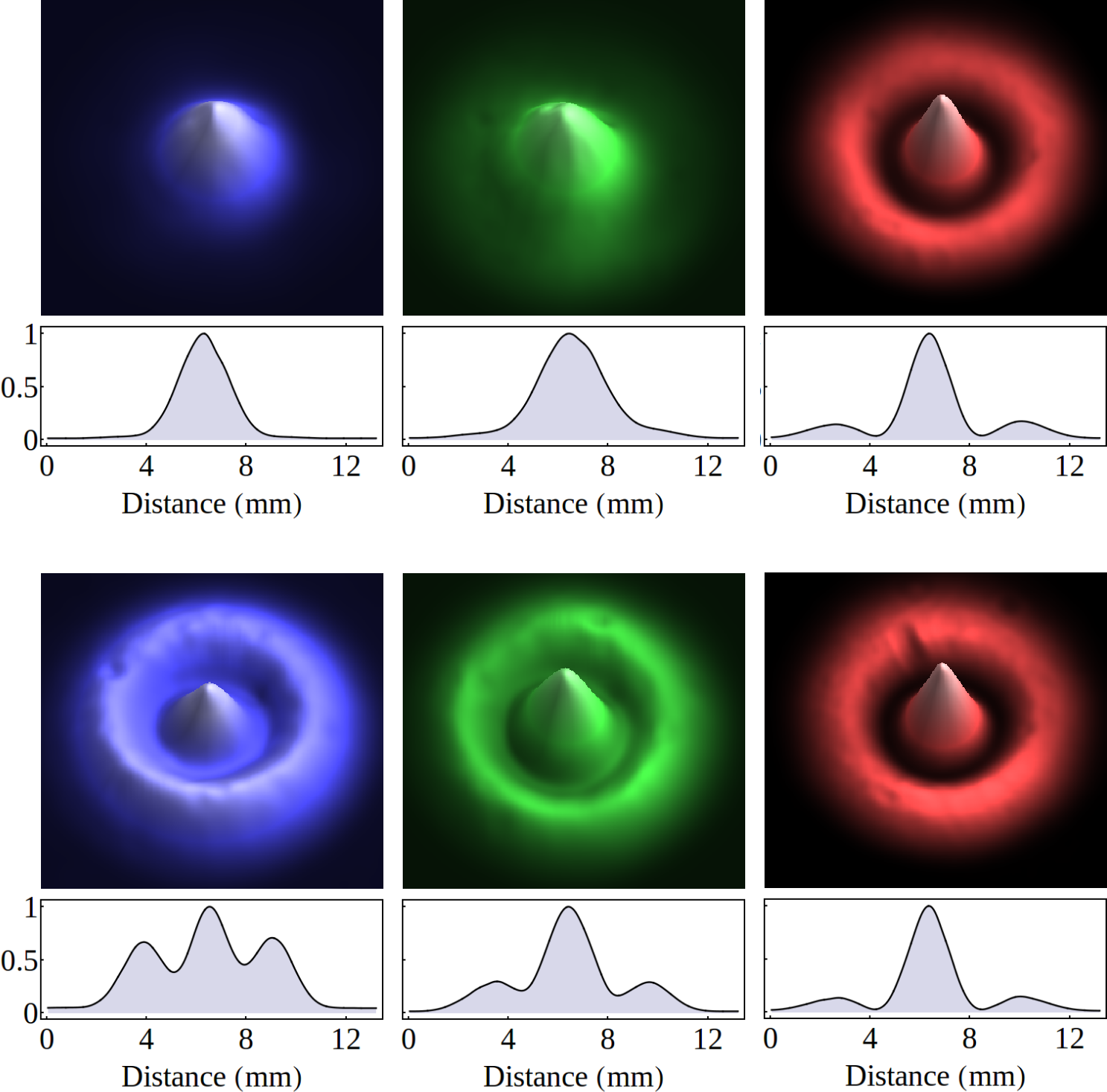}
\caption{Top and bottom rows show the two separate far-field mode profiles captured using a CCD camera beam profiler.
Each row from left to right indicates the beam profile of \as, pump, and \st. The top row shows a process that mainly supports FWM-\as, 
while the bottom row is a process that includes NLC-\as.}
\label{Fig:modes}
\end{figure}

The experiment is performed by launching a high peak power (pump) laser into the core of a short-length MMF. 
The laser is a 532~nm frequency doubled Nd:YAG with 500~Hz repetition rate, 650~ps pulse duration, and 
a peak power of approximately 10~kW. The fiber used in the experiment is Corning SMF-28$^\circledR$, which supports a few guided modes in the visible 
wavelength range. We recently showed in Ref.~\cite{2015-Mafi-Pourbeyram-OPEX} that the intermodal 
FWM generates a Stokes (\st, red color) and an antiStokes (\as, blue color) from vacuum fluctuations. We also confirmed that by coupling
the pump laser predominantly to the \lpzo~mode of the fiber, the main IM-FWM output \st-\as~peaks are generated 
at 656~nm (\lpzt~mode) and 447~nm (\lpzo~mode), respectively. ${\rm LP}$ stands for {\em Linearly Polarized} and 
follows the standard notation used for spatial modes in {\em Fiber Optics}~\cite{Okamoto}.
The measured spectrum containing the IM-FWM \st~and \as~sidebands and the measured mode profiles
are shown in Fig.~\ref{Fig:spectrum} and Fig.~\ref{Fig:modes}-top row, respectively.
The observed wavelengths and spatial modes of the \st-\as~pair agree
with theory: their wavelengths and spatial modes are determined using the conservation of energy and 
linear momentum (phase-matching) on the photons interacting via the IM-FWM process.
We have also observed that the above is not the only possible IM-FWM process: the energy and momentum conservation are also satisfied for a few other 
spatial mode combinations, resulting in {\em different} \st-\as~wavelengths~\cite{2016-Mafi-Pourbeyram-PRA,2015-Mafi-Pourbeyram-OPEX}. 
These other IM-FWM processes have also been 
observed in the measured output spectrum, albeit with a lower power because of their less efficient IM-FWM gain.
The \as~spectral feature of one such IM-FWM process is visible in Fig.~\ref{Fig:spectrum}, slightly red-shifted
from the dominant \as, while the corresponding \st~spectral feature is buried in the spectrometer noise.

{\em Of concern to this Letter}, we observed in some experiments that the dominant \as~at 447~nm appeared in \lpzt~instead of 
the phase-matched \lpzo~mode, while the dominant 656~nm \st~always appeared in \lpzt. Had the \lpzt~\as~appeared in a slightly different wavelength, 
we would have concluded that it belonged to another IM-FWM phase-matched process. However, the wavelengths of the commonly observed dominant \lpzo~\as~exactly 
matched that of the occasionally observed anomalous \lpzt~\as, and both processes seemed to be quite efficient. 
On the surface, this anomalous observation amounted to the appearance of an efficient non-phase-matched process, or in other words a 
violation of the conservation of momentum. The mode profiles of two separate observations for this puzzling behavior are shown in 
Fig.~\ref{Fig:modes}.
Using theory, we ruled out an accidental IM-FWM degeneracy. Another suspect was mode-coupling (MC): 
under linear MC the \lpzo~\as~could rotate to \lpzt~if the MMF were subject to certain perturbations. The occasional appearance of \lpzt~seemed to support 
the perturbation idea but the MMF was too short for microbending to play a role and the fiber was straight in the experiment. Moreover, the variations in 
the modal profile of the \as~along the fiber measured using the cut-back method did not conform to a linear MC behavior.
After further investigation primarily driven by theory and also noting in Fig.~\ref{Fig:modes} (bottom row) that this peculiar behavior
is always accompanied with a non-\lpzo~profile of the pump, we propose that this puzzling behavior is instigated by a peculiar combination of phase-matched 
parametric amplification (i.e. IM-FWM) and nonlinear mode coupling (NLC) that appear together in this experiment. {\em The rest of this Letter 
is to build a theoretical case and present the relevant experiments to verify this proposal}.

{\em Theory}: The experiment concerns with the degenerate-frequency FWM scheme, where (both) pump fields oscillate at same carrier frequency $\omega_p$
and couple to \st~and \as~at $\omega_s$ and $\omega_a$, respectively, via Kerr nonlinearity. 
The energy conservation is satisfied if $2\omega_p=\omega_s+\omega_a$. In order to construct a model to explain the experimental results, 
we assume that the pump field
propagates primarily in \lpzo~and \lpzt, and the former is assumed to carry considerably
more power. If the input Gaussian pump beam is perfectly mode matched to \lpzo~in the fiber, then no \lpzt~is 
excited. In practice, a slight (inevitable) mismatch results in the excitation
of \lpzt. For a simple defocusing, \lpzt~power can be shown to be always less than \lpzo\cite{2011-Mafi-OL}, while this may not be the 
case in the presence of a tilt or offset mismatch . 
This assumption is supported by the observed pump mode profiles in Fig.~\ref{Fig:modes}.
Including higher order modes only complicates the analysis but does not add to the essence of the intended treatment.
The IM-FWM process was shown in Ref.~\cite{2015-Mafi-Pourbeyram-OPEX} to be rooted in a degenerate-mode-pump process. \lpzo~carries the 
majority of the pump power; therefore, it is the mode which triggers the IM-FWM process.
This assumption is justified by the exponential dependence of the IM-FWM \st~and \as~powers on the pump power~\cite{2015-Mafi-Pourbeyram-OPEX}. 
The phase-matching equation for the IM-FWM process
can be written as $2\beta^{p}_{01}=\beta^{s}_{02}+\beta^{a}_{01}$, where $\beta^{p}_{01}$ indicates the propagation constant of the degenerate 
\lpzo~pump mode (subscripts indicate the spatial modes), and $\beta^{s}_{02}$ and $\beta^{a}_{01}$ are those of the \st~and \as, respectively.
Note that only the \lpzo~mode of \as~plays a role in the IM-FWM process. 

We propose that an additional NLC process triggered by a phase-matching of the form 
$\beta_{01}^{p}-\beta_{02}^{p}=\beta_{01}^{a}-\beta_{02}^{a}$ plays an important role in this fiber (this will be justified later). 
This relates to the beating between the two pump modes, which results in a Kerr-induced 
grating~\cite{2010-Fallnich-Andermahr,2014-Walbaum,2014-Fallnich-Hellwig}, hence coupling the two \as~\lpzo~and \lpzt~spatial modes.
In other words, we suggest that the \as~is generated via IM-FWM in the \lpzo~and is rotated to \lpzt~due to the nonlinear grating
formed by the beating between the two pump modes. Hereafter, we refer to the \lpzo~\as~as FWM-\as~and to the \lpzt~mode 
generated via the NLC at the \as~wavelength as NLC-\as.

The relevant processes discussed above can be read off directly from the Generalized Nonlinear Schr\"{o}dinger equation (GNLSE)
in Ref.~\cite{2012-Mafi} and be cast into a more simplified 
form, which includes only terms quadratic in the (strong) pump field: 
\begin{align}
\nonumber
\partial_z A^{a}_{01}&=i\mathcal{G}^{a}_{01}A^{a}_{01}+i(2\gamma_1 A^{a}_{02}A^{p}_{01}A^{p\ast}_{02} + \gamma_2 A^{p}_{01}A^{p}_{01} {A^{s\ast}_{02}}),\\
\nonumber
\partial_z A^{a}_{02}&=i\mathcal{G}^{a}_{02}A^{a}_{02}+i(2\gamma_3 A^{a}_{01}A^{p}_{02}A^{p\ast}_{01}),\\
\partial_z A^{s}_{02}&=i\mathcal{G}^{s}_{02}A^{s}_{02}+i (\gamma_4 A^{p}_{01}A^{p}_{01} A^{a\ast}_{01}).
\label{EQ:CE}
\end{align}

In these equations, $A^{a}_{01}$ represents the slowly varying envelope of \as~propagating in the spatial mode \lpzo~and so on. 
All $\gamma$'s are nonlinear coupling coefficients, which are determined as explained in 
Ref.~\cite{2012-Mafi} using the spatial mode profiles of the coupling fields at the specific frequencies and the third-order Kerr nonlinear
susceptibilities at the relevant frequencies. The terms multiplied by $\gamma_2$ and $\gamma_4$ are related to the IM-FWM process
and the terms multiplied by $\gamma_1$ and $\gamma_3$ govern the NLC process.
$\mathcal{G}$'s in the above equations describe the cross-phase-modulation (XPM) of the pump field on the \st, FWM-\as~and NLC-\as~and  
are defined as $\mathcal{G}^{j}_{l}\equiv2(\gamma^{jp}_{l,1}|A^p_{01}|^2+\gamma^{jp}_{l,2}|A^p_{02}|^2)$
where $j\in a,s$ and $l\in$~\lpzo, \lpzt. The propagation of the two pump modes can be reliably modeled by the undepleted-pump 
approximation and including only the (spatial) self-phase-modulation (SPM) and (spatial) XPM among the two pump modes~\cite{2013-Agrawal-Book}.

\begin{figure}[h]
\includegraphics[width=3.3in,]{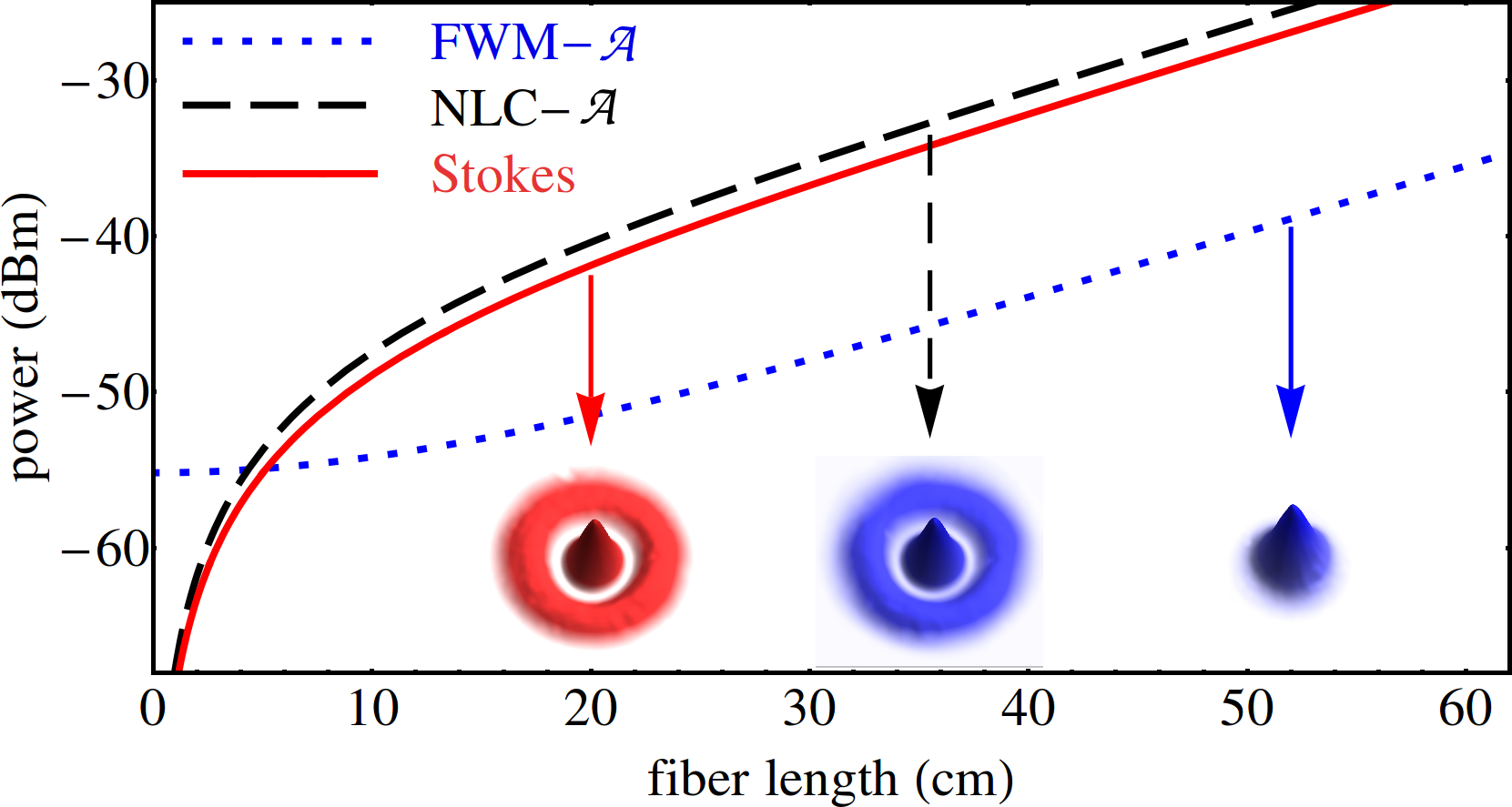}
\caption{Evolution of FWM Stokes, FWM-\as, and NLC-\as~ beams inside a short piece of multimode optical fiber
for $\mathcal{R}_p=20\%$.}
\label{Fig:theory}
\end{figure}

\begin{figure}[h]
\includegraphics[width=3.3in,]{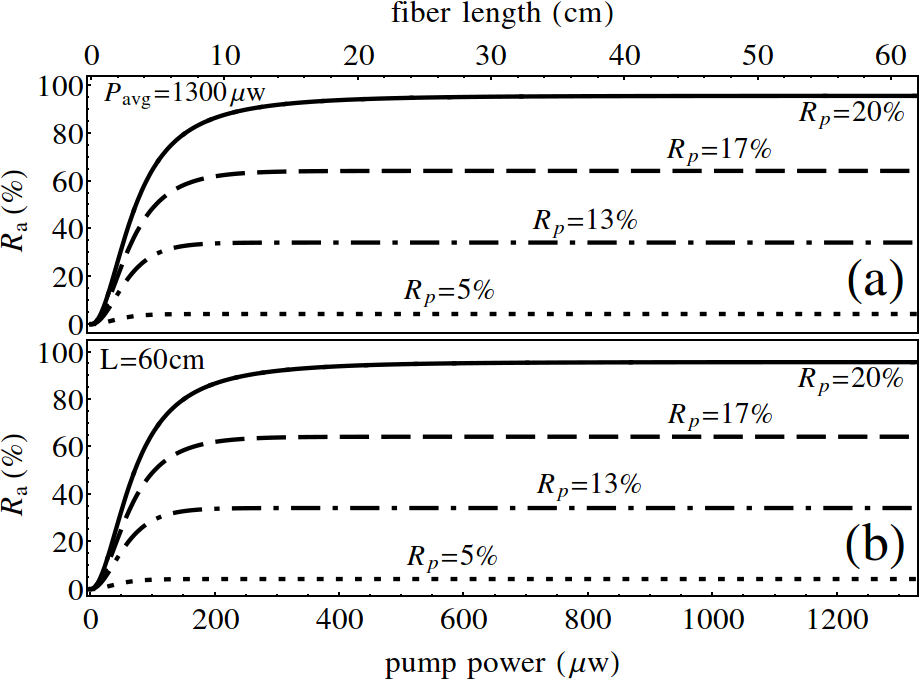}
\caption{(a) Modal conversion as a function of fiber length for a fixed pump power. (b) Modal conversion as a function of pump power for a fixed fiber length.}
\label{Fig:saturation}
\end{figure}
The simulations are carried out by solving Eqs.~\ref{EQ:CE}. The nonlinear coefficients are calculated as described above and 
in more detail in Refs.~\cite{2012-Mafi} and~\cite{2015-Mafi-Pourbeyram-OPEX}. The \st~and FWM-\as~are generated 
spontaneously, where we use one photon per transverse mode in FWM-\as~as the initial condition~\cite{1972-Smith}. 
Figure~\ref{Fig:theory} shows the growth of power for \st~(\lpzt, solid-red), FWM-\as~(\lpzo, dotted-blue), 
and NLC-\as~(\lpzt, dashed-black), as a function of the fiber length.
The simulations presented in Fig.~\ref{Fig:theory} show that the \st~and FWM-\as~powers grow with an exponential form 
asymptotically as expected. However, NLC between FWM-\as~and NLC-\as~results in a complex power exchange between
all three participating fields--the result is that the NLC-\as~power becomes considerably larger than FWM-\as~and becomes
comparable to \st~asymptotically. From an experimental observation point of view, this result may look quite puzzling because 
in the output Stokes appears primarily with an \as~which is not phase-matched and appears to violate 
the conservation of momentum. Our formalism shows that this behavior stems from the strong nonlinear interaction between the two 
pump modes, FWM generated \st~and FWM-\as, and their NLC to NLC-\as.

Figure~\ref{Fig:saturation}(a) shows the power ratio ($\mathcal{R}_{a}$) defined as\\
$\mathcal{R}_{a}$={\em NLC-\as/(total \as~power)} as a function of the propagation length. Different curves are plotted for different 
ratios $\mathcal{R}_p$={\em(LP$_{02}$~pump power)/(total pump power)}, for 1.3~mW average pump power. For $\mathcal{R}_p=0.05$, the pump power is predominantly in \lpzo; 
therefore, the efficiency of NLC is low and FWM-\as~dominates over NLC-\as. When $\mathcal{R}_p=0.20$, 
the NLC becomes quite efficient; therefore, NLC-\as~dominates over FWM-\as~in the output.
Figure~\ref{Fig:theory} is plotted for $\mathcal{R}_p=0.20$. The important observation in these plots is the saturation of the ratio of 
NLC-\as~to FWM-\as; therefore, the coupling between these two modes is not oscillatory and the experimental observer would detect 
the same ratio of the two spatial modes at the \as~wavelength when the fiber is cut-back for a sufficiently long optical fiber.
Note that as the fiber is cut-back, the FWM power drops exponentially with length; therefore, the curved rising regions in these
plots could not be measured in our setup.
Figure~\ref{Fig:saturation}(b) shows $\mathcal{R}_{a}$ plotted as a function of the (average) pump power for $L=60$~cm, 
where each curve is again plotted for a different value of $\mathcal{R}_p$. Similarly, the key observation here is that given a pump
coupling configuration, i.e. a fixed value of $\mathcal{R}_p$, the $\mathcal{R}_{a}$ increases with the pump power but eventually
saturates to a fixed value determined by $\mathcal{R}_p$. Again, only the saturated flat region can be measured in our experiment
due to the exponential dependence of the FWM process on the pump power.

{\em Experiment}: In order to obtain a quantitative measure of modal conversion in experiment, we have used a modal decomposition technique to calculate 
the weight of each spatial mode (FWM- versus NLC-\as) at the measured output beam. The same method was applied to assess
the weight of \lpzo~and \lpzt~modes for the pump and \st.  
The modal decomposition algorithm is similar to the one used in 
Ref.~\cite{2003-Fink-Skorobogatiy}: we calculate the spatial modes using a step-index profile and propagate them
using the ABCD transfer matrix method~\cite{Yariv-Book} to obtain the far-field intensity pattern at the beam profiler location. 
We then perform a least-square fitting to compare the measured total intensities between that of the beam profiler and the 
reconstructed beam. Modal weights and phases are the only two fitting parameters used in this analysis. We find that by only using the
two \lpzo~and \lpzt~modes it is possible to reconstruct the output beam accurately~\cite{pourbeyram2017saturable}.

\begin{figure}[htp]
\includegraphics[width=3.2in]{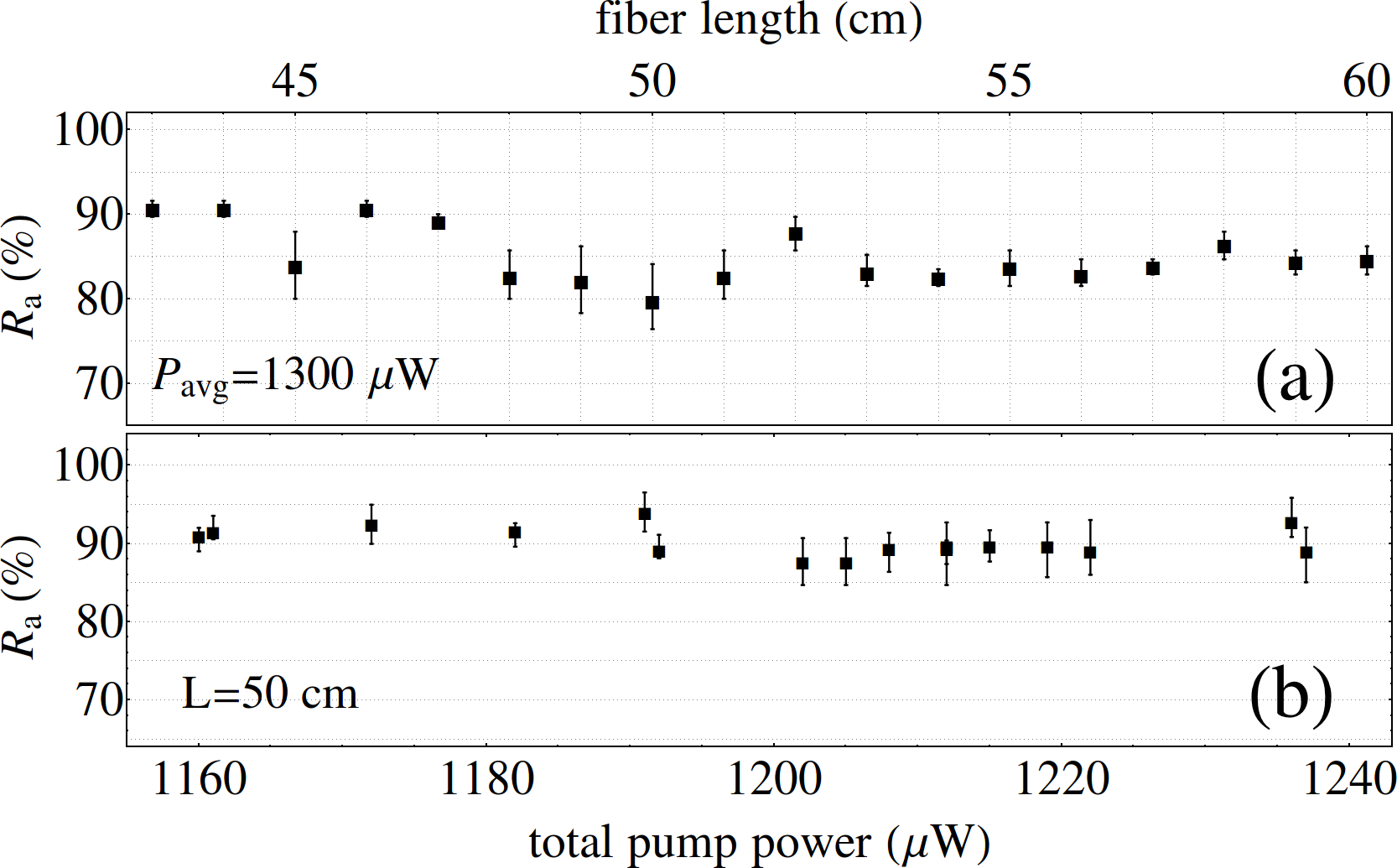}
\caption{Experimental measurements of modal conversion as function of the fiber length for a fixed pump power (a) 
and as a function of the total pump power for fixed fiber length (b).}
\label{Fig:ratio-exp}
\end{figure}
In Fig.~\ref{Fig:ratio-exp} (a), we show the NLC conversion efficiency of \as~($\mathcal{R}_{a}$) for a sample experiment conducted at 
1.3~mW average power as a function of length of the fiber. The fiber length started at 60~cm and was reduced by
the cut-back method. The error bars indicate the variation in the fitting of $\mathcal{R}_{a}$ when the least square integral
becomes twice the minimum value. Figure~\ref{Fig:ratio-exp}(a) demonstrates a saturation value in modal conversion of around 85$\pm$5$\%$,
where the conversion saturation is in agreement with the theoretical model presented here. Note that we explained earlier that 
the curved rising regions in theoretical plots of top right panel in Fig.~\ref{Fig:theory} could not be measured in our setup because of
their low power. In Fig.~\ref{Fig:ratio-exp} (b), we show $\mathcal{R}_{a}$ for a sample experiment conducted for a 
50~cm optical fiber as a function of the average pump power. Again, the NLC conversion efficiency shows the expected saturation 
in agreement with the theory.  

\begin{figure}[h]
\includegraphics[width=3.3 in]{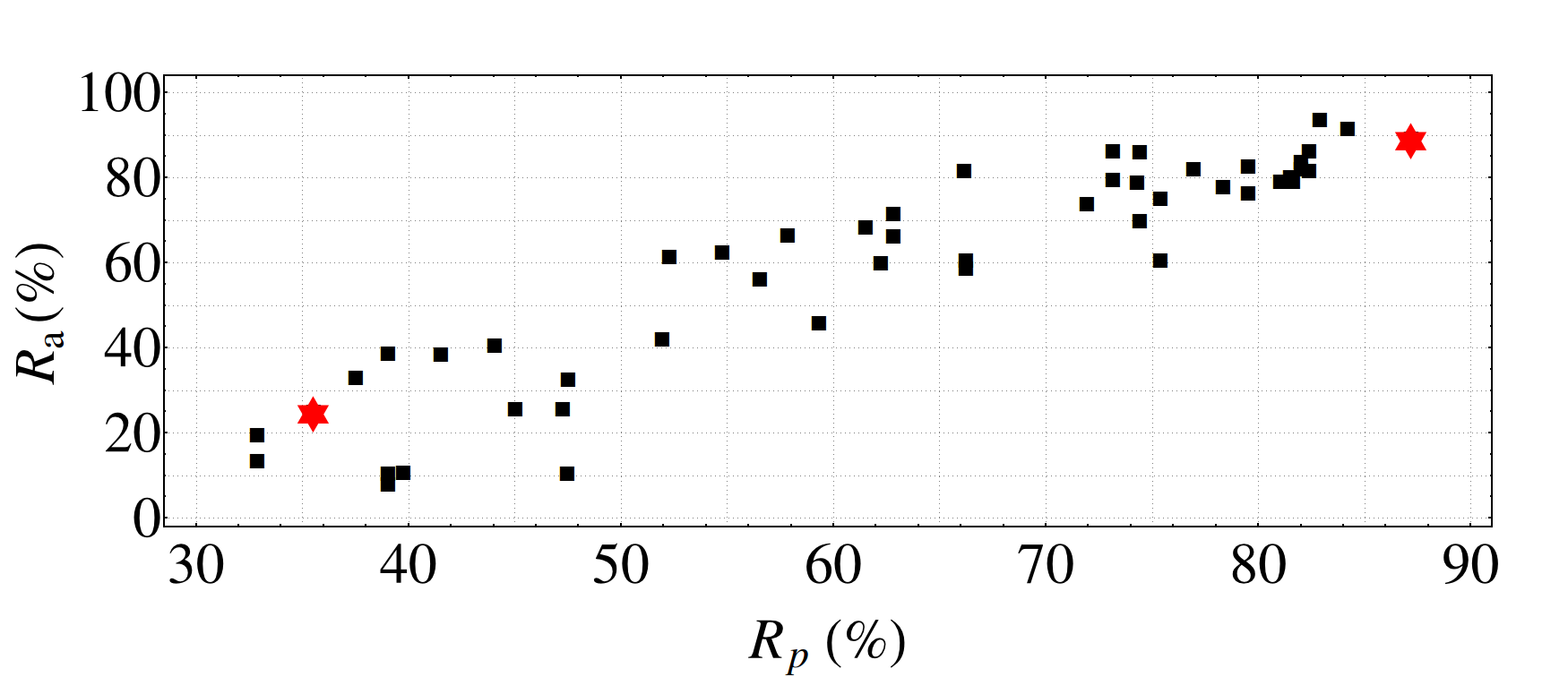}
\caption{Modal conversion as a function of $\mathcal{R}_p$ for 50 independent experiments. 
The left (right) red stars correspond to the top (bottom) rows in Fig.~\ref{Fig:modes}.}
\label{Fig:P-I}
\end{figure}

Another interesting observation is the correlation between $\mathcal{R}_{a}$ and $\mathcal{R}_p$. We note that $\mathcal{R}_p$
involves the measurements of the modal content of the pump. 
From Fig.~\ref{Fig:theory}, it is clear that for a sufficiently long fiber and in the presence of sufficient pump power,
$\mathcal{R}_{a}$ saturates to above 90\% level if $\mathcal{R}_p>$20\%. In Fig.~\ref{Fig:P-I}, we show a scatter plot of
this correlation for 50 different measurements for various fiber lengths, powers, and pump couplings. The scatter plot clearly
shows a positive correlation between $\mathcal{R}_{a}$ and $\mathcal{R}_p$. However, the data indicates that there is no 
maximum value for $\mathcal{R}_p$ to have an active NLC process. We would like to point out that the maximum limit for $\mathcal{R}_p$ 
at the level presented in Fig.~\ref{Fig:theory} is an artifact of several assumptions made 
in the (minimalistic) theoretical model in Eq.~\ref{EQ:CE}. The maximum level can be easily modified by 
relaxing the assumption of {\em perfect} phase-matching of the form $\beta_{01}^{p}-\beta_{02}^{p}=\beta_{01}^{a}-\beta_{02}^{a}$,
which results in a phase-mismatch term in $\gamma_1$ and $\gamma_3$ terms in Eq.~\ref{EQ:CE}. In addition, 
the relative values of $\gamma$'s and the total pump power have substantial impact the saturation behavior. 

There is another possible phase-matched term that we did not discuss so far: by subtracting the (approximate) NLC phase-matching equation
from the FWM phase-matching equation, we find $\beta^{p}_{01}+\beta^{p}_{02}=\beta^{a}_{02}+\beta^{s}_{02}.$
This is a non-degenerate pump mode FWM process. In the following, we justify why this term has been ignored 
in our analysis by showing that it represents a highly non-efficient process. Essentially, such non-degenerate pump mode processes
have been shown to favor FWM with much smaller spectral shifts~\cite{1982-Stolen-Bjorkholm} and are highly inefficient at such 
large spectral shift values explored here. We note that there are no {\em a priori} guarantees for proper phase-matching for the NLC of
the two \as~modes. In fact, for the configuration presented here, we have verified that it is not exactly phase-matched;
rather, a small shift of around 20~pm in wavelength of NLC-\as~(corresponding to $\beta^{a}_{02}$) is required 
for perfect phase-matching. However, 20~pm is well within the phase-matching bandwidth of the main FWM process, 
so the NLC process is quite efficient. However, a similar exercise to properly phase-match the above-mentioned
non-degenerate pump mode FWM requires a wavelength shift of 4~nm in \as~and 7~nm is \st~wavelengths, which
is well beyond the efficient FWM bandwidth of the main process.

Another important point is the possibility of temporal walk-off between the two pump modes~\cite{2014-Walbaum}. 
The group-velocity difference between the \lpzo~and \lpzt~pump modes at 532~nm wavelength is 
$\Delta\nu_g=15.7\times10^4$~m/s, resulting in only 2.3~ps temporal separation over 60~cm, which is negligible 
considering the 680~ps duration of pump pulses. The coherence length of the FWM process presented here is also calculated to be 
$l_c\approx38$~cm. Therefore, the fiber lengths used in our experiments are comparable to $l_c$, making the the FWM process efficient.
The only point of consideration is our continuous-wave (CW) assumption in solving the GNLSE because the 680~ps pulse duration
translates to a 14~cm long pulse in the fiber, which is 4 times shorter than the length of the fiber. However, none of the results 
presented in this paper are affected by this approximation, which was adopted for simplicity, because we are not making any direct 
quantitative comparison between observed experiments and numerical simulations. 

In conclusion, the presented results provide an explanation for the appearance of linear momentum non-conservation in
nonlinear optical experiments involving MMFs. 
In turn, this can be viewed as a scheme for deliberate excitation of desired spatial modes
with specific amplitude weights, which remain robust because of the saturable conversion behavior explained above.
In comparison, grating-based methods for mode conversion are length-dependent and often require accurate length tuning of the
fiber, while the proposed method suggests a modal conversion independent of the fiber length and pump power.
The scheme presented may be utilized in scenarios where a special modal content is needed to observe a nonlinear phenomenon, such as
the generation of multimode solitons reported in Refs.~\cite{2016-Wise-Zhu,2015-Wise-Wright-OPEX}, or for an all optical switch~\cite{2014-Walbaum}.
The strong coupling between the FWM and NLC processes is quite intriguing, and we expect
similar behavior may be possible with modes of non-zero angular momentum, where the NLC may be responsible for 
the unexplained report on the violation of angular momentum conservation~\cite{1981-Hill}. We emphasize that the investigations reported here
mainly concern cases with relatively short lengths of fiber, similar to~\cite{2015-Wise-wright,2012-Mafi}, 
while for longer optical fibers other important 
factors such as linear mode coupling may play more dominant roles~\cite{2014-Agrawal-Xiao}. From an application standpoint, 
the key observations are fourfold: first, unlike the linear coupling, which results in a periodic 
mode conversion and requires careful tuning of the fiber length to obtain a specific mode, the output saturates 
here to a particular mode for a sufficiently long fiber; second, the modal conversion can be very efficient
and as high as 90\%; third, the saturated conversion efficiency can be tuned by controlling the spatial
coupling of the pump into the fiber; and fourth, the saturated mode can be amplified via the IM-FWM process.
The measured conversion efficiency is comparable or even higher than the earlier reported ones 
using an optically induced long-period grating~\cite{2010-Fallnich-Andermahr,2016-Fallnich-Schnack,2013-Fallnich-Hellwig,2014-Fallnich-Hellwig}.

Last but not least, using the scheme presented here one can generate a spatial Bell-state~\cite{Kovlakov} via 
the FWM process in a MMF without transverse mode subspace post selection.
In the spontaneous FWM limit the biphoton component of the \st-\as~state is dominant and 
using a dichroic mirror, one can project the state to a single photon in the \as~wavelength.
For a case that the conversion efficiency is 50\% there is a 50-50 chance for a photon in the 
\lpzo~mode to convert to \lpzt. In this case the \as~photon will be in a spatial Bell-state
of the form
\begin{align}
\ket{\psi_{\rm a}}\propto
\ket{1}_{01}\ket{0}_{02}+\ket{0}_{01}\ket{1}_{02}=\ket{\Psi}^+,
\end{align}
where the subscripts indicate the spatial modes. We emphasize that the power exchange 
happens at the single-photon level for the interaction between the spontaneous FWM and NLC processes.

The authors acknowledge support by the National Science Foundation (NSF) Grant No. 1522933.

\bibliography{blist}

\end{document}